# The new line of attack of analyses of NMR in 2D structures.

G. M. Minkov★, S. A. Negashev★, O. E. Rut★, A.V. Germanenko★ ,
O. I. Khrykin✸, V. I. Shashkin✸ and V. M. Danil'tsev ✸
★*Institute of Physics and Applied Mathematics, Ural State University*
*620083 Ekaterinburg, Russia*
✸*Institute for Physics of Microstructures of RSA, 603600 N. Novgorod, Russia*

There was shown that Fourier transform of the negative magnetoresistance (NMR) which is due to interference correction to the conductivity contains the information about the area distribution function of the closed paths and about area dependence of the mean length of closed paths $\overline{L}(S)$. Based on this line of attack we suggest the method of analysis of NMR and use it for data treatment of the NMR in 2D structure with doped barrier. There was shown that in structure investigated $\overline{L}(S)$ dependence is determined by the scattering anisotropy.

The anomalous magnetoresistance which is observed at low temperature in "dirty" metals and semiconductors was provided by adequate explanation after creating the theory of the quantum correction to the conductivity [1,2,3]. The interference correction to the conductivity gives the main contribution to the NMR in 2D structures at low temperatures. This correction arises from interference of the electron waves scattered along closed paths in opposite directions. In magnetic field perpendicular to the 2D layer this interference is destroyed because of the phase shift between the correspondent amplitudes and thus leads to NMR.

The existing theory of weak-field magnetoresistance is developed under condition $k_F l >> 1$ ($k_F$ is the Fermi wave vector, $l$ is the mean free path). In this case the quasiclassical treatment is valid and the interference correction to the conductivity can be written as sum of the contributions from closed paths [4]

$$\Delta\sigma = 2\pi l^2 G_0 \sum_i W_i \exp\left( -\frac{L_i}{l_\varphi} \right) \tag{1}$$

where $G_0 = \dfrac{e^2}{2\pi^2\hbar}$ , $W_i$ is the probability density to find the electron in the starting point after passing on i-th trajectory with length $L_i$, $l_\varphi = v_F\tau_\varphi$, $v_F$ is the velocity at Fermi energy, $\tau_\varphi$ is the phase breaking time, the factor $\exp(-L_i / l_\varphi)$ is taking into account the probability of the breaking phase at passing on i-th trajectory.

For calculation of the magnetic field dependence of NMR the sum (1) is represented as sum of the contributions of trajectories with different number of collisions N [1,5,6]

$$\Delta\sigma(B) \equiv \sigma(B) - \sigma(0) = 2\pi l^2 G_0 \sum_N \sum_i W_i^N \exp\left(-\frac{L_i}{l_\varphi}\right)\left(1-\cos\left(\frac{2\pi S_i^N B}{\Phi_0}\right)\right) \tag{2}$$

where где $S_i^N$ -is the area of i-th trajectory with N collisions, $\Phi_0 = 2\pi c\hbar / e$ is the elementary flux quantum, $\left(1-\cos(2\pi S_i^N B/\Phi_0)\right)$ is the factor taking into account the destruction of interference by magnetic field. The diagrammatic technique [1,6] gives possibility to calculate the sum (2) and gives the analytical result for NMR when two conditions are to be met: (i) At random scatters distribution, and (ii) at isotropic scattering which corresponds to the scattering by short range potential. At diffusion approximation ( when the collisions number for actual trajectory is large (mach greater then unity)) this gives [1]

$$\begin{gathered} \Delta\sigma(B) = aG_0 F(B,\tau_\varphi,\tau_p) \\ F(B,\tau_\varphi,\tau_p) = \Psi(0.5 + \frac{B_{tr}\tau_p}{B\tau_\varphi}) - \ln\frac{B_{tr}\tau_p}{B\tau_\varphi} \end{gathered} \tag{3}$$

where $\Psi(x)$ is the logarithmic derivation of $\Gamma$-functhion, $\tau_p$ is the elastic scattering time, $B_{tr} = \dfrac{\hbar c}{2el^2}$. ( Without taking into account the electron-electron interaction $a = 1$.) Beyond the diffusion limit the function $F(B,\tau_\varphi,\tau_p)$ was calculated in [6].

Just the expression (3) is used for analysis of the experimental data of NMR and more or less agrees with magnetic field dependence of NMR. It gives possibility to determine the phase breaking time and its temperature dependence. At this way of analysis of the experimental data the reasons of the some deviation of NMR from (3) remain unclear. They may be connected with some correlation in scatters distribution, scattering anisotropy and so on.

We suggest the method for the analysis of NMR which gives some information about the statistic of closed paths and its dependence on the scatter distribution and peculiarities in scattering. To make clear the essence of this method let us rewrite (1) decomposing $W_i$ as a sum of contributions of paths with given area

$$\Delta\sigma(B) = 2\pi l^2 G_0 \sum_S W(S) \exp\left(-\frac{\overline{L}}{l_\varphi}\right)\left(1-\cos\left(\frac{2\pi SB}{\Phi_0}\right)\right) \tag{4}$$

where $W(S) = \sum_i W_i^S$ is the distribution function of closed paths with area, $W_i^S$ is the length distribution function of the trajectories with given area $S$, and we introduce $\overline{L} = \overline{L}(S, l_\varphi)$ so that

$$\exp\left(-\frac{\overline{L}(S,l_\varphi)}{l_\varphi}\right)=\frac{\sum_i W_i^S \exp\left(-\frac{l_i^S}{l_\varphi}\right)}{W(S)} \qquad (5)$$

From (4) one can see that Fourier transform of NMR is equal

$$\Phi(S)=\sqrt{\frac{\pi}{2}}2\pi l^2 G_0 W(S)\exp\left(-\frac{\overline{L}(S,l_\varphi)}{l_\varphi}\right) \qquad (6)$$

Thus the Fourier transform of NMR contains the information about the distribution function of closed path with area $W(S)$, and about $S$ and $l_\varphi$ dependence of $\overline{L}$. For further analysis we propose that $\overline{L}=\overline{L}(S,l_\varphi)=S^\alpha f(l_\varphi)$. The numerical calculations of the function $\overline{L}(S,l_\varphi)$ ( these calculations will be published elsewhere) and analysis of expression (3) show that this assumption is valid in wide range of $S$, $l_\varphi$ and for isotropic scattering $\alpha \approx 0.67$. At first sight, it seems that the mean length of closed paths is proportional to square root of the area S, but taking into account that the trajectories with larger area are more "winding" one can conclude this fact leads to more strong area dependence of $\overline{L}$. The ionised impurity scattering is the main mechanism of the momentum relaxation at low temperature. For this scattering mechanism the scattering is anisotropic, the small angle scattering gives the main contribution. It is clear that for strong anisotropy the all trajectory became close to the rings and therefore $\overline{L}\propto S^{0.5}$, i.e. $\alpha=0.5$. Thus, value of $\alpha$ is determined by scattering anisotropy.

To determine experimentally the value of $\alpha$ one can measure $\Delta\sigma(B)$ at two temperatures, i.e. at different $l_\varphi$, then find

$$A(S)\equiv \ln\left[\frac{\Phi(S,T_1)}{\Phi(S,T_2)}\right]=S^\alpha\,(f(l_\varphi^2)-f(l_\varphi^2)) \qquad (7)$$

and from $A$-vs-$S$ dependence determine $\alpha$.

Let us analyse the experimental results. We have measured the NMR in heterostructure 0.3 μm $n^-$-GaAs /50 A-$In_{0.07}Ga_{0.93}$ As/0.3 μm $n^-$GaAs. δ-doping by Si layers was arranged on both sides of the well for distance of 100 A. The measurements in wide range of magnetic field (up to 6 T) and temperature (1.5-40 K) show that in the structure investigated only one size-quantized 2D subband is occupied and the main contribution to the conductivity comes from electrons in $In_{0.07}GaAs_{0.93}$ quantum well with density $n=2.5\times10^{11}$ $cm^{-2}$ and mobility $\mu=1.1\times10^4$ $cm^2/V$ sec.

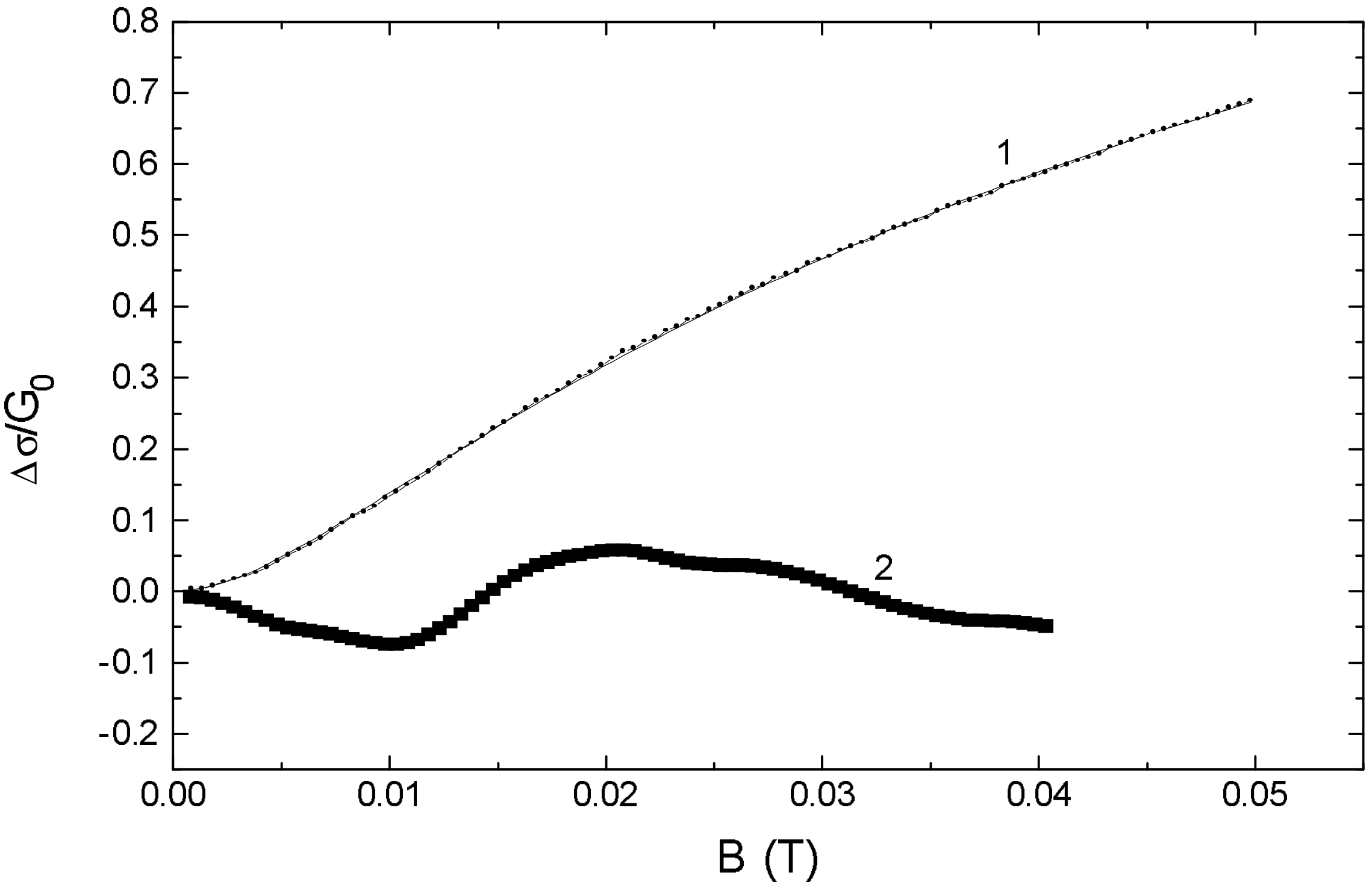

Fig.1. (1)The magnetic field dependence of $\Delta\sigma(B)/G_0$ at T=1.5 K. The points are the experimental data, the curve is theoretical dependence (3) at $a = 0.7$ и $\tau_\varphi = 5.4\times10^{-12}$ sec (2). The difference between theoretical and experimental dependences $(\Delta\sigma(B)_{theor} - \Delta\sigma(B)_{\exp})\times 20$

The magnetic field dependence of NMR is shown in Fig.1. Usually the expression (3) is used to analyse the MNR with $a$ and $\tau_\varphi$ as fitting parameters. The solid curve in Fig.1 was obtained by just the same method and at the first sight agrees well with experimental data at magnetic field $B < B_{tr} \approx 0.038$ T at $a = 0.7$ и $\tau_\varphi = 5.4\times10^{-12}$ sec. But a closer look shows the difference between theory and experimental data (curve 2 in Fig.1). This difference leads to the fact that the parameters $a$ and $\tau_\varphi$ vary in range 0.65-0.88 и $(4.4-6)\times10^{-12}$,respectively, at fitting in different interval $B$ inside the range $0 < B < B_{tr}$. Thus, the accuracy of determination the values $a$ and $\tau_\varphi$ is 15-20%. Most of authors associate less than unite value of the prefactor $a$ with contribution of electron-electron interaction ( Maki-Tompson term) [7]. But, from our point of view in structure investigated it results from invalidity of diffusion approximation $\tau_\varphi/\tau_p >> 1$. Really, in our case $\frac{\tau_\varphi}{\tau_p} \approx 5-10$, and as it is clear from analysis of the calculations NMR beyond diffusion approximation [6], for such ratio of $\tau_\varphi$ and $\tau_p$ , NMR well described by the expression (3) also, but with prefactor less then unity.

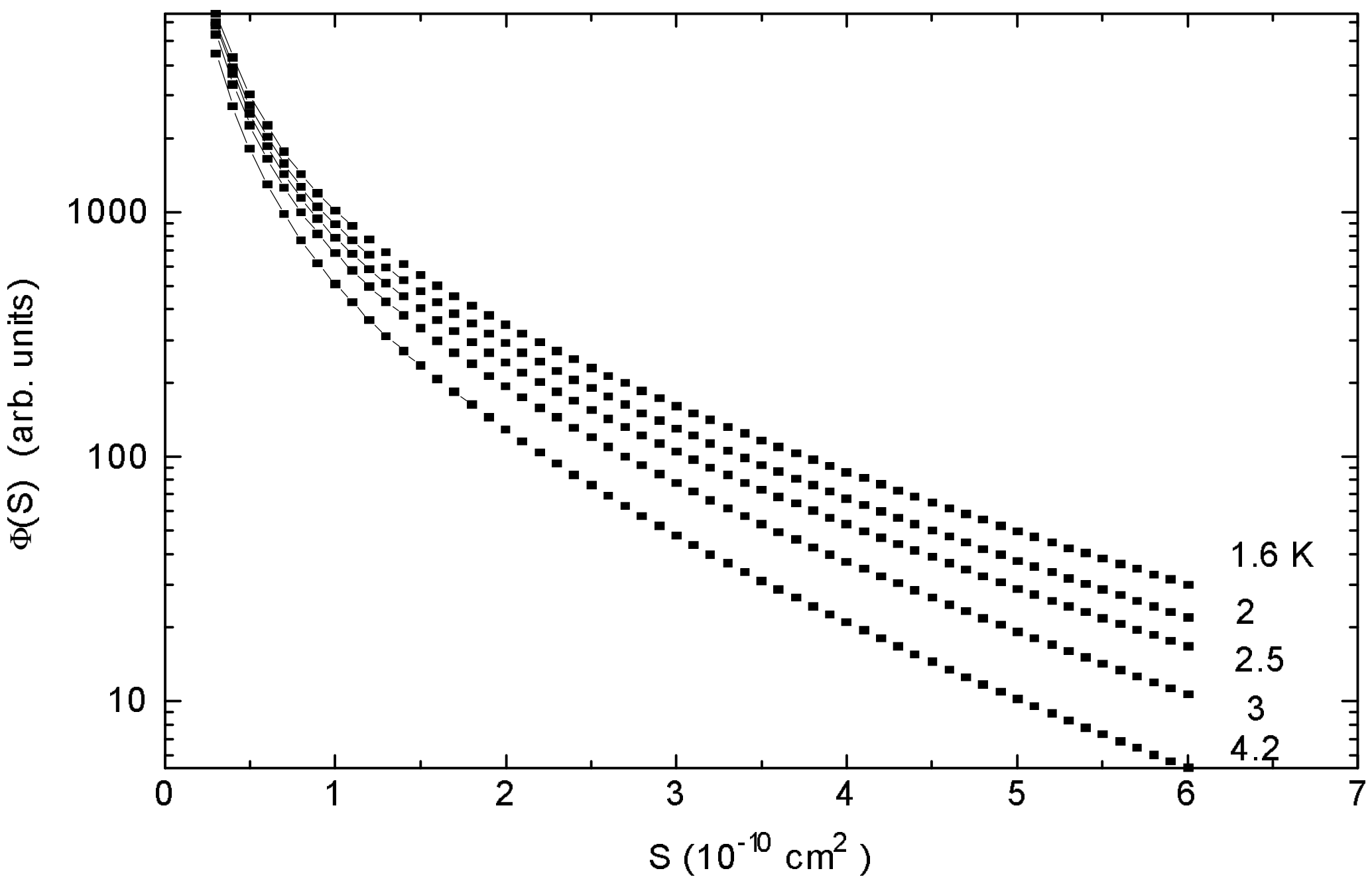

Fig.2. The Fourier transforms of the experimental data at different temperatures.

Let us consider what provides the method put forward above. In Fig 2 we present the Fourie transforms of $\Delta\sigma(B)$ at different temperatures. As it is clear from (7), the plot $\ln(A(S))$-vs-$\ln(S)$ gives the value of $\alpha$. We have obtained $\alpha = 0.52 \pm 0.05$. To illustrate the accuracy in determination $\alpha$, $A(S)$-vs-$S^{-\alpha}$ dependencies for different $\alpha$ are plotted in Fig. 3.

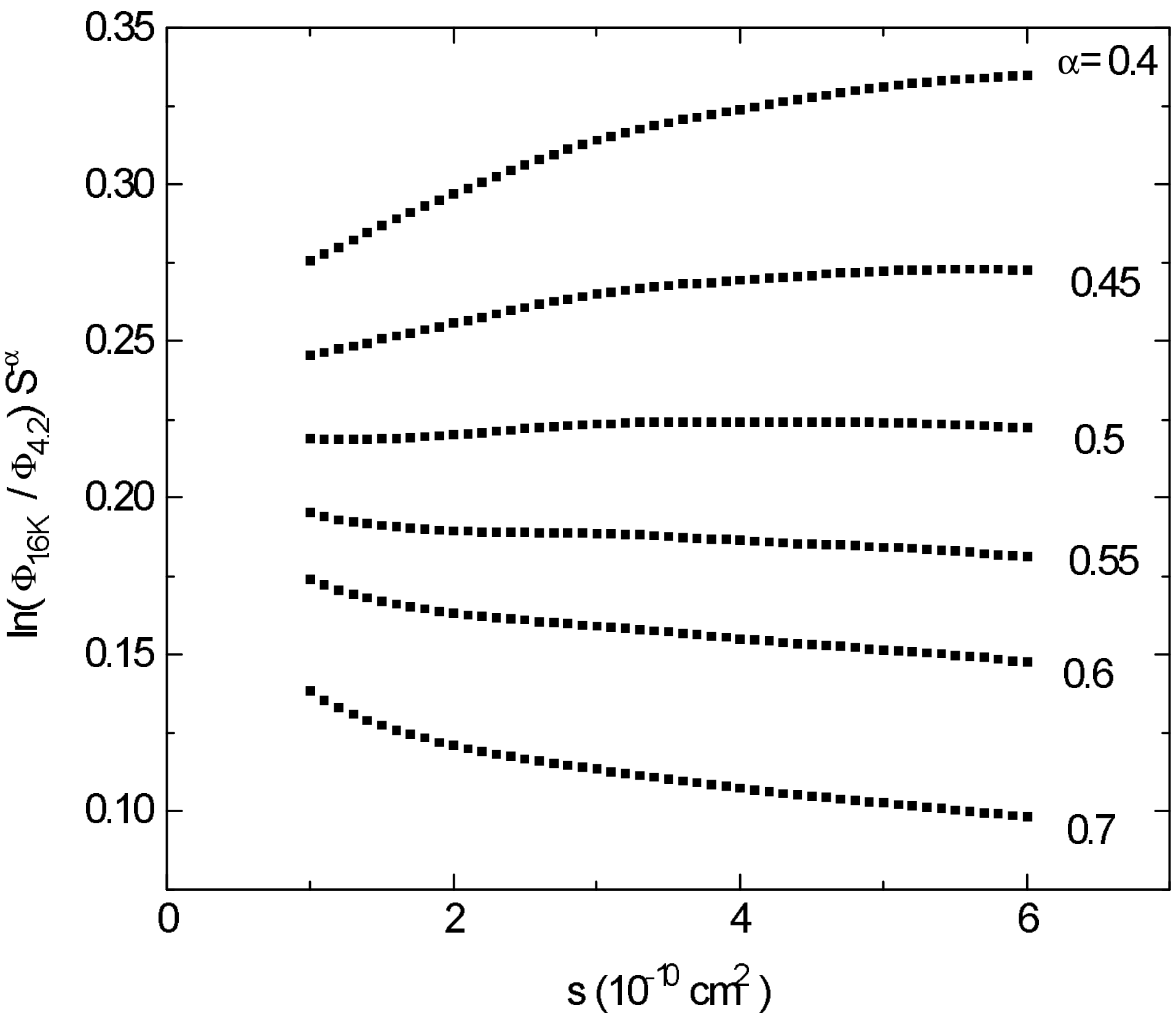

Fig.3. The dependences of $A(S)S^{-\alpha} = \ln(\Phi_{1.5K}(S)) - \ln(\Phi_{4.2K}(S))$ - vs- $S$ for different $\alpha$ .

As mentioned above, for isotropic scattering (i.e. for scattering by short range potential) the value of $\alpha$ is equal to 0.67. To ensure that this is correct one can analyse the NMR given by (3) using the method discussed above. It gives: i - assumption that $\overline{L}(S,l_\varphi) = S^\alpha f(l_\varphi)$ is valid in wide range of $S$ и $l_\varphi$; ii - $\alpha$ =0.67.

We believe that the less then 0.67 value of $\alpha$ in structure investigated is the result of the scattering anisotropy. Really, the main mechanism of the momentum relaxation in our condition is the scattering by the potential of ionised impurities of δ-layers which is smooth in quantum well. The estimations show that probability of small angle scattering is about 15-20 times lager than backscattering probability.

In principle, the method put forward in this paper gives possibility to determine the distribution function of closed path with area *W(S)* . Really, because the length $l_\varphi$ tends to infinity when T→0, the extrapolation Φ(S,T) to T=0 gives the area distribution function W(S) (see (6)). But, for such extrapolation it is necessary to have the measurements of $\Delta\sigma(B)$ for lower temperatures.

In conclusion, the method of the analysis of NMR put forward in this communication provides a way to obtain information on statistic of closed paths and in this way on the scattering anisotropy, correlation of impurity distribution and so on.